\documentclass[%
reprint,
superscriptaddress,
amsmath,amssymb,
aps,
pra,
]{revtex4-2}

\usepackage{mathrsfs}
\usepackage{graphicx}
\usepackage{dcolumn}
\usepackage{bm}
\usepackage{amsmath, amssymb}
\usepackage{makecell}
\usepackage{multirow}
\usepackage{float}
\usepackage{color}
\usepackage{stmaryrd}
\usepackage{array}
\usepackage{setspace}
\usepackage{mathtools}
\usepackage{booktabs}

\makeatletter

\newcommand{\Rmnum}[1]{\expandafter\@slowromancap\romannumeral #1@}
\makeatother                                                                   

\begin{document}                                                                
\preprint{APS/123-QED}
	
\title{Topological Leading Angular Structure around Bound States in the Continuum} 
	
\author{Nan Zhang}
\email{nan25@vt.edu}
\affiliation{Bradley Department of Electrical and Computer Engineering, Virginia Tech, Blacksburg, Virginia 24061, USA}
\affiliation{Department of Mathematics, City University of Hong Kong, Kowloon, Hong Kong, China}
\author{Ya Yan Lu}
\affiliation{Department of Mathematics, City University of Hong Kong, Kowloon, Hong Kong, China}
	
\date{\today}
	
\begin{abstract}
We show that the angular structure of the leading radiation around a bound state in the continuum
(BIC) directly encodes its topology. 
This angular information can be determined from the BIC itself
and used to design prescribed topological polarization textures.
As examples, we identify BICs carrying high topological charges
$q=4$ and $q=-5$ in $C_{6v}$ photonic crystal slabs,
addressing the question of whether high-charge BICs exist and how they can be systematically realized.
We further show that zeros of the leading radiation can drive both
topological transitions and the bifurcation of new BIC families.
These findings may enable new approaches to polarization engineering in structured light and resonant photonics.
\end{abstract}

\maketitle

Bound states in the continuum (BICs) are nonradiating eigenstates embedded
in the radiation continuum~\cite{Neumann1929PZ,Friedrich85PRA,Marinica08PRL,
	Plotnik11PRL,Hsu13Nature,Hsu16NRM,Kivshar2023PU}.
Their infinity quality ($Q$) factors have been widely exploited to enhance
light-matter interactions, with applications in lasing~\cite{Mocella2015PRB,Kodigala17Nature,Yuri21NC,Rong23NM,do2025room,Cui25NP,farazi2026bound}, sensing~\cite{Kivshar19NP,Azzam2021,nan2025angular}, and
quantum optics~\cite{Maria22Science,Rivera23PNAS}.
Since the seminal work of Zhen {\em et al.} in 2014~\cite{Zhen14PRL},
the topological properties of BICs have also attracted broad interest~\cite{Zhen14PRL,Bulgakov17PRATopo,JZi18PRL,Alu18NP,bulgakov2017topological,JZi19PRL,Zhen19Nature,JZi20NP,Yoda20PRL,kostyukov2022ring,ochiai2024generation,Zi22PRLSpin,Kang25PRL,Dong26PRL}.
A BIC is a far-field polarization singularity in momentum space,
associated with a vortex carrying an integer topological
charge $q$, enabling applications such as the generation of optical vortex beams~\cite{JZi20NP,QSong20Science}, topological lasing~\cite{Zi25PRLTopLasing,arjas2024high},
and, more recently, momentum-space spin textures~\cite{Guo20PRL,Rao25PRL}.
To date, numerous experimental and numerical studies have investigated BIC topology
through the polarization maps of resonant states in momentum space,
revealing diverse topological configurations and transitions involving
the merging, splitting, and annihilation of BICs as structural parameters
vary~\cite{Bulgakov17PRATopo,JZi18PRL,Alu18NP,bulgakov2017topological,JZi19PRL,Zhen19Nature,
	JZi20NP,Yoda20PRL,Kang21PRL,Kang22LSA,Kang25PRL,Dong26PRL}.

In this Letter, we show that the angular structure of the leading radiation around a BIC 
provides a unified description of its topology and topological transitions. 
This structure can be determined from the BIC itself, 
without requiring information about nearby resonant states, 
thereby providing a systematic route for designing prescribed topological structures. 
Since the topological characterization of BICs was introduced in 2014~\cite{Zhen14PRL}, 
the reported topological charges of BICs in periodic dielectric structures have, 
to our knowledge, remained within $|q|\leq 3$~\cite{Zhen14PRL,Bulgakov17PRATopo,JZi18PRL,Alu18NP,JZi19PRL,Zhen19Nature,JZi20NP,Yoda20PRL,hu2022global,Bogdanov22PRB,Kang22LSA,su2026high}. 
Our theory addresses two fundamental questions:
whether BICs with higher topological charges can exist
and how such states can be systematically realized.
Here, we identify BICs with $q=4$ and $q=-5$ in $C_{6v}$ photonic crystal (PhC) slabs.
We also show that the appearance of the leading radiation zeros marks
the critical conditions for topological transitions and the bifurcation
of new BIC families under structural variation. Isolated angular zeros,
in particular, give rise to sharp variations in the momentum-space
polarization and unusual bifurcation behavior.
Our results provide a framework for uncovering and engineering BIC topological structures, 
with implications for structured light and resonant photonics.

We begin by introducing the leading radiation around a BIC. 
We consider a structure that is periodic in the $xy$ plane and finite in $z$. 
Only the zeroth diffraction order is assumed to be open in each of the 
upper and lower half-spaces. 
For ease of presentation, we focus on nondegenerate at-$\Gamma$ BICs with 
up-down mirror symmetry $\sigma_z$ and $C_n$ rotational symmetry, 
$n=3,4,6$. 
More general cases are discussed in the Supplemental Material~\cite{SM}. 
We parameterize nearby resonant states by writing the Bloch wavevector as
${\bm\kappa}=\delta{\bm\theta}/L$, where $\delta>0$ is a small parameter,
$L$ is a characteristic length, and
${\bm\theta}=(\cos\theta,\sin\theta)$, with $-\pi\leq\theta<\pi$,
specifies the angular direction in momentum space.
The far-field radiation vector ${\bm d}({\bm\kappa})$ admits an expansion in
$\delta$, with ${\bm d}_j(\theta)$ denoting its $j$-th coefficient.
These coefficients can be determined directly from the BIC. 
Taking the BIC as the unperturbed state,
we expand the resonant state in $\delta$.
Substitution into the Maxwell equations yields a hierarchy of equations
that determines the field corrections order by order~\cite{Zhang25PRL,SM}.
The outgoing far-field amplitudes of the $j$-th field correction then
give ${\bm d}_j(\theta)$.
We define $\mu$ as the smallest order for which 
${\bm d}_\mu(\theta)$ is not identically zero in $\theta$, so that 
\begin{equation} 
	{\bm d}({\bm \kappa}) 
	\sim 
	\delta^\mu{\bm d}_\mu(\theta). 
	\label{eq:dasymptotic} 
\end{equation} 
The angular zeros of the leading radiation are collected in 
${\cal Z}_\mu=\left\{\theta:{\bm d}_\mu(\theta)=0\right\}$. 
As shown in Refs.~\cite{Zhang25PRL,SM}, energy conservation gives 
$Q\sim Q_{2\mu}(\theta)/\delta^{2\mu}$ for $\theta\notin{\cal Z}_\mu$, 
whereas higher-order scaling can occur along directions in ${\cal Z}_\mu$.

Let ${\bm S}$ denote the $2\times 2$ reduced scattering matrix at the BIC frequency.
For a lossless and reciprocal structure, ${\bm S}$ is symmetric and unitary.
Moreover, $C_n$ rotational symmetry with $n=3,4,6$ requires
${\bm S}=s{\bm I}$ with $|s|=1$.
If the BIC field is scaled to be time-reversal symmetric,
the leading-order radiation coefficient strictly satisfies~\cite{SM}
\begin{equation}
	{\bm d}_\mu(\theta)
	=
	(-1)^{\mu+1}s\overline{{\bm d}_\mu}(\theta).
\end{equation}
Although the radiation vector ${\bm d}({\bm \kappa})$ is generally complex, 
after a fixed phase transformation, ${\bm d}_\mu(\theta)$ can be made real.
Consequently, the radiation near an at-$\Gamma$ BIC with $C_n$,
${\cal T}$, and $\sigma_z$ symmetries is nearly linearly polarized.
Breaking any one of these symmetries can lift the reality constraint,
allowing elliptically polarized leading radiation~\cite{Zhao24PRL,Qin23LSA}.
The case with only $C_2$ symmetry is discussed in the Supplemental Material~\cite{SM}.

\begin{figure*}[htbp]
	\centering
	\includegraphics[scale=0.33]{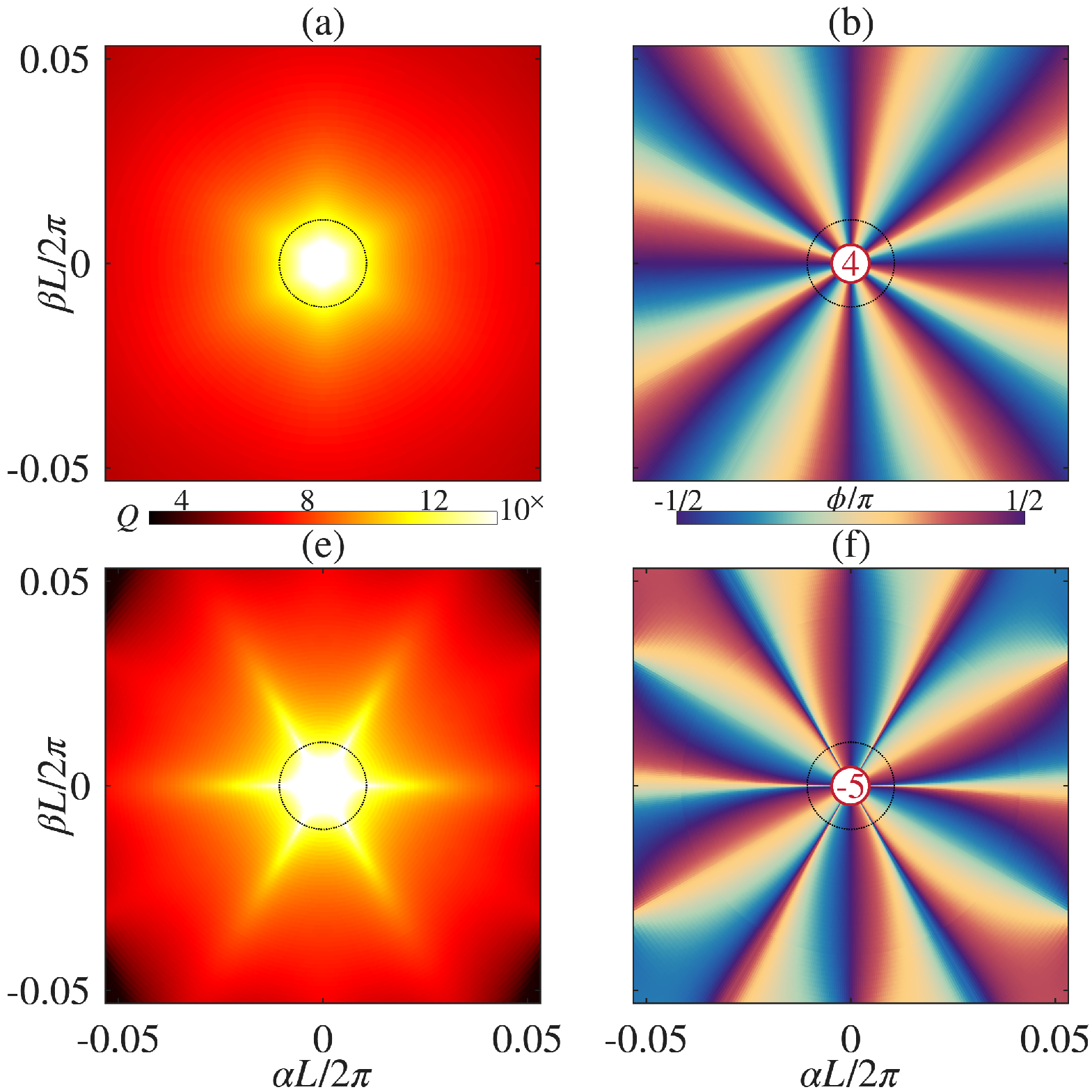}
	\includegraphics[scale=0.33]{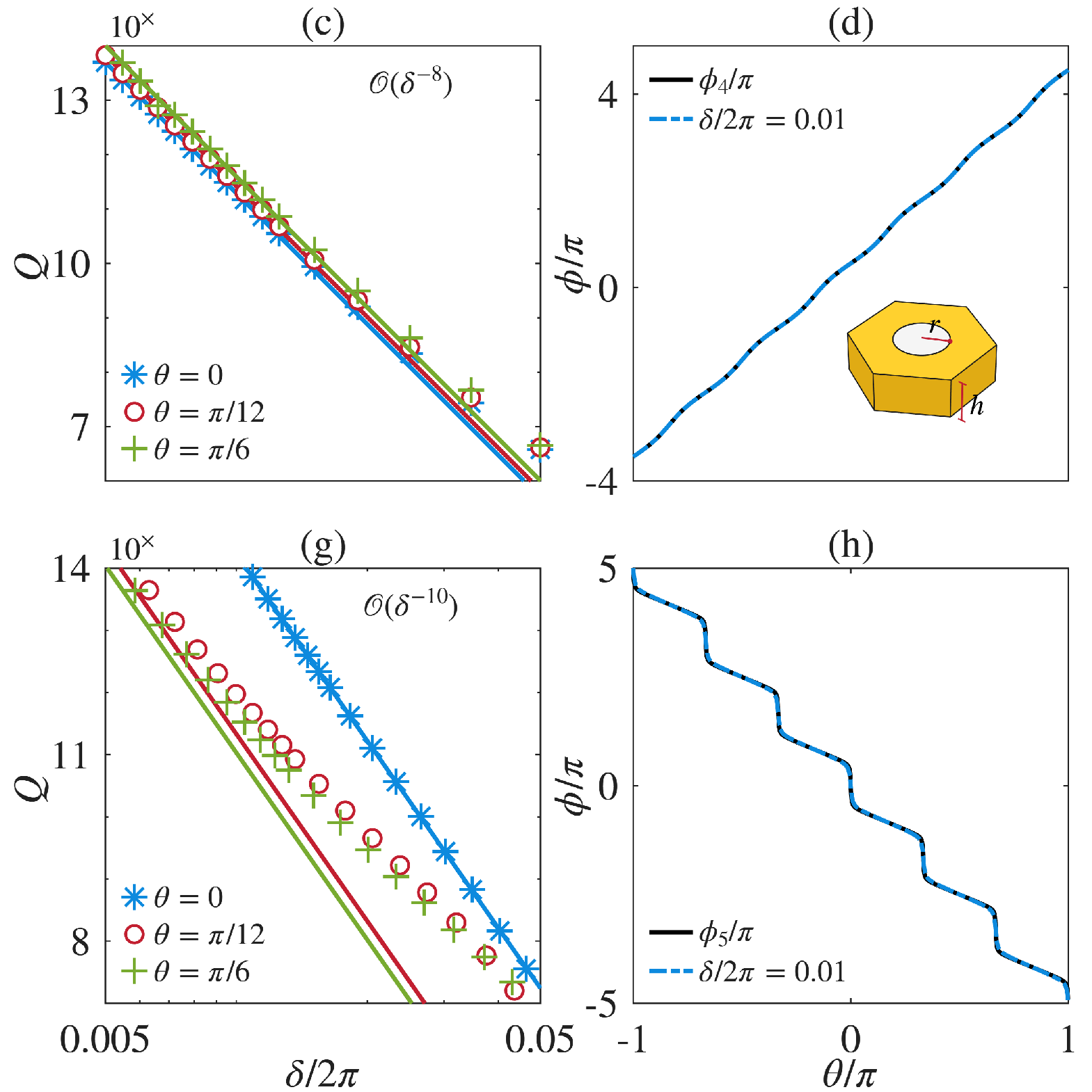}
	\caption{
		High-topological-charge at-$\Gamma$ BICs. A unit cell is shown in the inset of panel (d).
		(a)--(d) A BIC with $\mu=4$ and $q=4$.
		(e)--(h) A BIC with $\mu=5$ and $q=-5$.
		(a,e) $Q$ in the ${\bm \kappa}$ plane, where ${\bm \kappa}=(\alpha,\beta)$.
		(b,f) Polarization angle $\phi/\pi$ of the nearby resonant states.
		The dashed circles correspond to $\delta/2\pi=0.01$.
		(c,g) $Q$ as a function of $\delta$ along $\theta=0$, $\pi/12$, and $\pi/6$.
		Markers denote direct numerical results and solid lines show the asymptotic behaviors
		$Q\sim Q_8(\theta)/\delta^8$ and
		$Q\sim Q_{10}(\theta)/\delta^{10}$, respectively.
		(d,h) Unwrapped polarization angle predicted by the leading radiation compared
		with that of the resonant state at $\delta/2\pi=0.01$.
	}
	\label{fig:q4_qminus5}
\end{figure*}

We next study the angular structure of ${\bm d}_\mu(\theta)$. 
Since ${\bm d}({\bm \kappa})$ is generally analytic in ${\bm\kappa}$,
${\bm d}_\mu(\theta)$ is a homogeneous polynomial of degree $\mu$
in $\cos\theta$ and $\sin\theta$.
Its angular dependence is therefore restricted to the finite set of harmonics 
\begin{equation} 
	{\cal M}_\mu 
	= 
	\{-\mu,-\mu+2,\ldots,\mu-2,\mu\}. 
\end{equation} 
It can then be written as 
\begin{equation} 
	{\bm d}_\mu(\theta) 
	= 
	\sum_{m\in{\cal M}_\mu} 
	{\bm R}_{m\theta}{\bm c}_{\mu,m},
	\label{eq:vector_angular_decomposition} 
\end{equation} 
where ${\bm c}_{\mu, m}$ are constant vectors
and ${\bm R}_{m\theta}$ is the rotation matrix through an angle ${m\theta}$.
For a BIC in a $C_n$-symmetric structure,
let $\varphi=2\pi/n$ and let $\tau$ denote its rotation eigenvalue.
Rotational symmetry further requires
\begin{equation}
	{\bm d}_\mu(\theta+\varphi)=\tau{\bm R}_\varphi{\bm d}_\mu(\theta).
\end{equation}
Substituting this relation into Eq.~\eqref{eq:vector_angular_decomposition},
we find that a nonzero coefficient ${\bm c}_{\mu,m}$ is allowed only when
\begin{align}
	&\tau=1:
	\quad m\equiv1\pmod n,
	\label{eq:m_selection_plus}\\
	&\tau=-1:
	\quad m\equiv1+n/{2}\pmod n,
	\label{eq:m_selection_minus}
\end{align}
where the second line applies only for even $n$.
We use ${\cal M}_{\mu}^{n,\tau}$ to denote the set of 
angular harmonics allowed at order $\mu$ for a BIC in the representation $\tau$ of a $C_n$ structure.
For the lowest few orders $\mu$, the sets ${\cal M}_{\mu}^{n,\tau}$ are 
listed in Table~\ref{tab:allowed_harmonics}.
\begin{table}[htbp]
	\caption{Sets ${\cal M}_{\mu}^{n,\tau}$ of allowed angular harmonics for the lowest
		few orders $\mu$.}
	\label{tab:allowed_harmonics}
	\centering
	\begin{tabular}{c|c|c|c|c|c}
		\toprule
		& $C_3$ & \multicolumn{2}{c|}{$C_4$} & \multicolumn{2}{c}{$C_6$}\\
		$\mu$
		& $\tau=1$
		& $\tau=1$
		& $\tau=-1$
		& $\tau=1$
		& $\tau=-1$\\
		\midrule
		$1$ & $1$      & $1$       & $-1$       & $1$      & --\\
		$2$ & $-2$     & --        & --         & --       & $-2$\\
		$3$ & $1$      & $-3,1$    & $-1,3$     & $1$      & --\\
		$4$ & $-2,4$   & --        & --         & --       & $-2,4$\\
		$5$ & $-5,1$   & $-3,1,5$  & $-5,-1,3$  & $-5,1$   & --\\
		$6$ & $-2,4$   & --        & --         & --       & $-2,4$\\
		\bottomrule
	\end{tabular}
\end{table}

We relate the leading angular radiation to the topological charge of a BIC.
For a closed contour ${\cal C}$ in momentum space enclosing the BIC and no
other polarization singularities, the topological charge is defined by
\begin{equation}
	q
	=
	\frac{1}{2\pi}
	\oint_{\cal C}
	\nabla_{\bm\kappa}\phi
	\cdot {\rm d}{\bm\kappa},
	\label{eq:BIC_charge}
\end{equation}
where
$\phi=0.5\arg({\cal S}_1+i{\cal S}_2)$
is the polarization angle of the nearby resonant states, and
${\cal S}_1$ and ${\cal S}_2$ are the Stokes parameters.
To connect this definition with the leading angular radiation, we take
${\cal C}={\cal C}_\delta$ to be a sufficiently small circle around the BIC.
Since ${\bm d}_\mu(\theta)$ is real, we
represent it by the complex scalar
\begin{equation}
	f_\mu(\theta)
	=
	d_{\mu x}(\theta)+i d_{\mu y}(\theta)
	=
	\sum_{m\in{\cal M}_\mu^{n,\tau}}
	a_{\mu,m}e^{im\theta},
	\label{eq:complex_angular_decomposition}
\end{equation}
where $a_{\mu,m}$ are constant coefficients.
Equation~\eqref{eq:dasymptotic} and the reality of
${\bm d}_\mu(\theta)$ give
\begin{equation}
	{\cal S}_1
	+i{\cal S}_2
	\sim
	\delta^{2\mu}f_\mu^2(\theta).
	\label{eq:stokes_fmu}
\end{equation}
If $f_\mu(\theta)\ne0$ for all $\theta$, ${\cal Z}_\mu =\varnothing$, then
${\cal S}_1+i{\cal S}_2$ remains nonzero on ${\cal C}_\delta$ for
sufficiently small $\delta$, and its winding is determined by
the leading term in Eq.~\eqref{eq:stokes_fmu}.
Therefore,
\begin{equation}
	q=\operatorname{wind}f_\mu.
\end{equation}
Thus, although the topological charge is defined from the polarization of nearby resonant states, 
it is encoded in the
angular structure of the leading radiation.
If any of the symmetries assumed above is broken, 
${\bm d}_\mu$ need not be
real, and Eq.~\eqref{eq:stokes_fmu} no longer holds in this form.
The general case is discussed in the Supplemental Material~\cite{SM}.

If ${\cal M}_\mu^{n,\tau}$ contains only one harmonic $m$, then
$q=m$. If multiple harmonics are allowed, the 
topological charge can be obtained by directly evaluating $\operatorname{wind}f_\mu$.
In particular, for the two-harmonic case with 
$m_-<m_+$, the winding is determined by the harmonic with the larger 
coefficient magnitude:
\begin{equation} 
	q 
	= 
	\begin{cases} 
		m_-, 
		& 
		|a_{\mu,m_-}| 
		> 
		|a_{\mu,m_+}|,\\[3pt] 
		m_+, 
		& 
		|a_{\mu,m_+}| 
		> 
		|a_{\mu,m_-}|. 
	\end{cases} 
	\label{eq:two_harmonic_charge} 
\end{equation} 
The equality $|a_{\mu,m_-}|=|a_{\mu,m_+}|$ is the corresponding
transition condition, at which $f_\mu=0$ along some critical direction
$\theta_c$. More generally, for any number of allowed harmonics, the
angular zeros can be computed directly.

Our theory also separates the physical origins of the $Q$-factor scaling and the topological charge.
Energy conservation determines the asymptotic $Q$-factor scaling through the radiation order $\mu$,
whereas symmetry constrains the angular structure of ${\bm d}_\mu$,
which determines $q$.
Importantly, the topological charge is generally not equal to the radiation order.
In many cases, $|q|\neq\mu$.
Suppressing lower-order radiation to increase $\mu$
does not necessarily lead to a BIC with a large topological charge.
For example, for $C_{6}$ with $\tau=-1$, $\mu$ can reach $6$,
whereas the maximum topological charge satisfies $|q|=4$.
Moreover, for $C_3$, increasing $\mu$ from $5$ to $6$ can actually
reduce the maximum attainable magnitude of the topological charge.
The equality $|q|=\mu$ occurs only for particular angular structures.

We demonstrate the validity and utility of the theory through two examples of high-charge BICs.
High topological charges can support higher-order polarization vortices and
richer momentum-space polarization textures, making them particularly
relevant to structured-light applications~\cite{JZi20NP,Rao25PRL}.
Here, we further impose an in-plane mirror symmetry $y\mapsto-y$ and
consider $C_{nv}$ structures.
The mirror symmetry reduces each allowed angular-harmonic vector
${\bm c}_{\mu,m}$ to a single independent real coefficient.
The number of independent real conditions required for all lower-order
radiation terms to vanish defines the codimension of a BIC of order $\mu$.
Our theory predicts that the largest
charge magnitude attainable at codimension one is $|q|=4$, realized by
a $C_{6v}$ BIC with $\tau=-1$ and $\mu=4$ when
$|a_{4,4}|>|a_{4,-2}|$, yielding $q=4$.
At codimension two, the largest attainable magnitude increases to
$|q|=5$, realized by a $C_{6v}$ BIC with $\tau=1$ and $\mu=5$ when
$|a_{5,-5}|>|a_{5,1}|$, yielding $q=-5$.

Such high-charge BICs are not straightforward to identify by conventional searches.
Their realization requires not only the suppression of lower-order radiation,
but also appropriate relations among the angular harmonics of the first
nonzero higher-order radiation.
This makes direct searches over momentum and structural parameter spaces
challenging.
Our formulation instead extracts the relevant angular information directly
from the BIC through ${\bm d}_\mu$, thereby providing a systematic route to
targeting prescribed high-charge BICs.
Specifically, ${\bm d}_\mu$ is obtained from the far-field radiation of the
$\mu$-th correction field in the wavevector expansion, calculated through a
sequence of source-driven Maxwell equations evaluated at the BIC frequency~
\cite{Zhang25PRL,Zhang26Fourier,SM}.

We directly search for these two BICs in a dielectric slab
of refractive index $3.5$, embedded in air and patterned by a triangular 
lattice of circular air holes with lattice constant $L$. 
For $q=4$, we fix the radius at $r=0.358L$ and vary the height $h$ to satisfy the required condition, obtaining a {\em B}$_2$ BIC at $h=0.961L$ and $\omega_*L/2\pi=0.709$.
For $q=-5$, we vary both $r$ and $h$ to satisfy the corresponding conditions, obtaining an {\em A}$_1$ BIC at $r=0.191L$, $h=0.991L$, and $\omega_*L/2\pi=0.707$.
Details of the numerical search are given in the Supplemental Material~\cite{SM}.

Figure~\ref{fig:q4_qminus5} verifies the predicted leading orders and 
topological charges. 
Writing ${\bm\kappa}=(\alpha,\beta)$, 
Figs.~\ref{fig:q4_qminus5}(a) and \ref{fig:q4_qminus5}(e) show the 
$Q$ factors near the two BICs, while 
Figs.~\ref{fig:q4_qminus5}(c) and \ref{fig:q4_qminus5}(g) verify the 
asymptotic scaling along representative directions. 
The {\em B}$_2$ BIC follows 
$Q\sim Q_8(\theta)/\delta^8$, whereas the {\em A}$_1$ BIC follows 
$Q\sim Q_{10}(\theta)/\delta^{10}$. 
Unlike previously reported high-order scaling confined to selected 
momentum directions, the $1/\delta^{10}$ scaling here holds in all 
directions, with only the prefactor $Q_{10}(\theta)$ depending on 
$\theta$~\cite{Yuan20PRAPert,Luo23PRA,Liu2024OE,Zhang2025Super}.

The corresponding polarization textures are shown in
Figs.~\ref{fig:q4_qminus5}(b) and \ref{fig:q4_qminus5}(f).
Along the dashed circle with $\delta/2\pi=0.01$, the polarization angle
winds by $8\pi$ and $-10\pi$, giving $q=4$ and $q=-5$, respectively.
Figures~\ref{fig:q4_qminus5}(d) and \ref{fig:q4_qminus5}(h) compare
$\phi$ obtained from the resonant states with
$\phi_4=\arg f_4(\theta)$ and $\phi_5=\arg f_5(\theta)$ predicted by the
leading radiation.
Their agreement confirms that ${\bm d}_\mu$ captures the polarization
winding and hence the BIC topology.

We further relate the leading angular structure to topological
transitions under structural variation. For a given resonant band, the
BICs form a set in ${\cal P}\times{\rm BZ}$, where ${\cal P}$ denotes the
structural-parameter space and ${\rm BZ}$ the Brillouin zone. A
one-parameter structural path ${\bm p}={\bm p}(\eta)$ defines a slice. 
Along such a path, a critical structure can occur
when the leading angular radiation develops a zero. 
At such critical points, new BIC branches can bifurcate, while the
topological configuration can change across them. 
A rigorous, general order-by-order construction of the local BIC branches
is given in the Supplemental Material~\cite{SM}. Here, we retain only the
leading-order description of these transitions.
We consider two basic types of critical structures. In the first, the
leading radiation order remains unchanged, but an isolated angular zero
appears at $\eta=0$. In the second, the leading radiation order increases
at $\eta=0$. More
complicated cases can be treated similarly.

We begin with the first case, in which ${\bm d}_\mu$ contains multiple
angular harmonics. For the two-harmonic case in
Eq.~\eqref{eq:two_harmonic_charge}, the transition occurs at
${\bm p}_*={\bm p}(0)$ when
$|a_{\mu,m_-}|=|a_{\mu,m_+}|$, so that ${\bm d}_\mu$ vanishes along
an isolated angular direction $\theta_c$. Near $(\theta_c,{\bm p}_*)$,
the radiation has the local form
\begin{equation}
	{\bm d}({\bm\kappa};\eta)
	\sim
	\delta^\mu
	\left[
	\eta{\bm a}
	+
	(\theta-\theta_c){\bm b}
	+
	\delta^p{\bm c}
	\right],
	\label{eq:fixed_mu_local}
\end{equation}
where
${\bm a}=\partial_\eta{\bm d}_\mu(\theta_c;0)$,
${\bm b}=\partial_\theta{\bm d}_\mu(\theta_c;0)$, and $p$ is the
smallest positive integer such that
${\bm c}={\bm d}_{\mu+p}(\theta_c;0)\ne0$.
This local form explains the sharp polarization variation near $\theta_c$ at $\eta=0$. 
In this case, the angular and higher-order radiation terms
become comparable when
$|\theta-\theta_c|\sim\delta^p$. If
${\bm b}\not\parallel{\bm c}$, the variation of their relative weights
produces an $O(1)$ change in polarization within an angular layer of
width $O(\delta^p)$. For $\eta\ne0$, nearby off-$\Gamma$ BICs
bifurcating from the critical point are determined, to leading order,
by the vanishing of the bracketed term in
Eq.~\eqref{eq:fixed_mu_local}, thereby ensuring conservation of the
total topological charge.
\begin{figure}[htbp]
	\centering
	\includegraphics[scale=0.33]{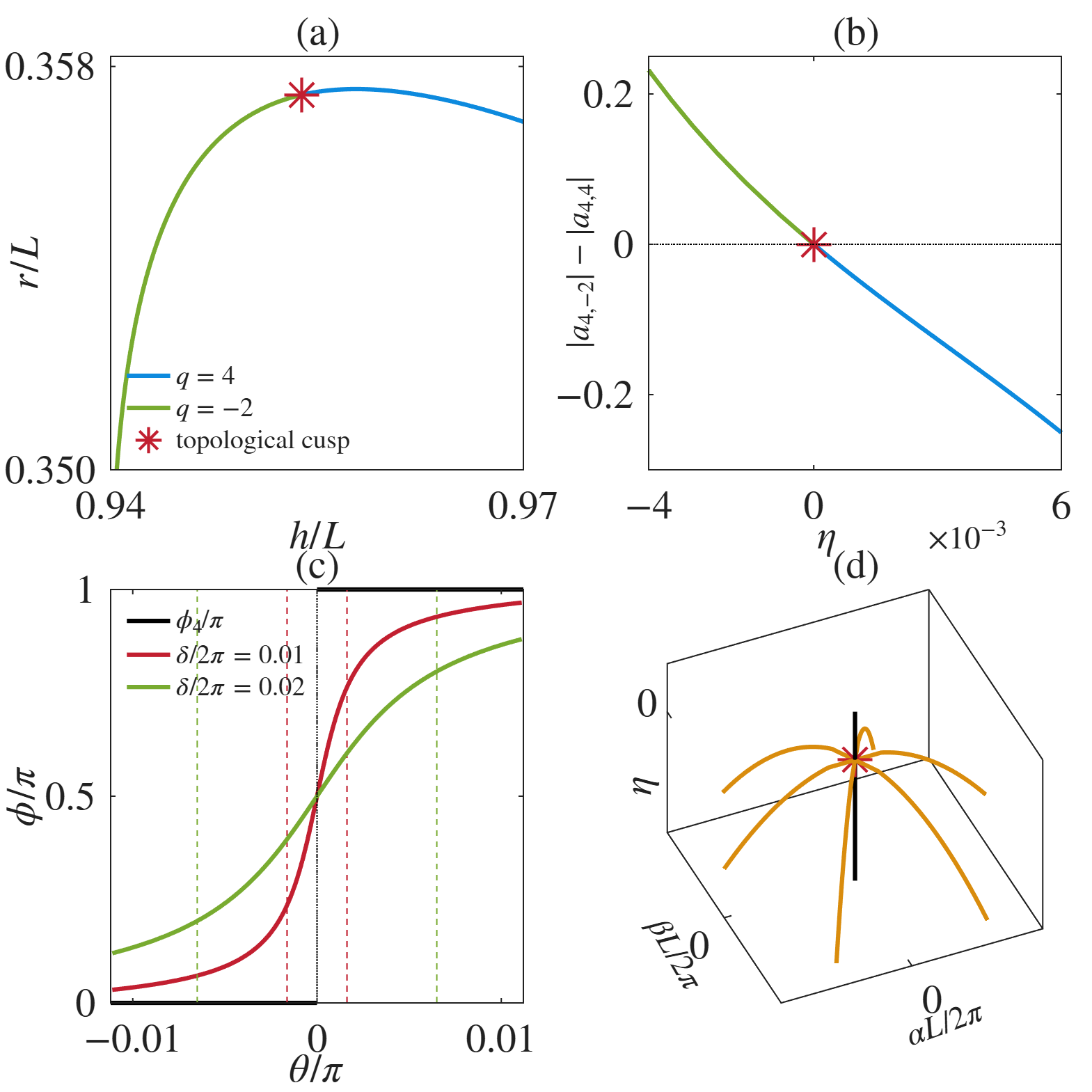}
	\caption{
		Topological transition along a family of $\mu=4$ BICs.
		(a) The family of BICs in the $(h,r)$ parameter plane.
		The green and blue branches correspond to $q=-2$ and $q=4$,
		respectively, and the red asterisk marks the cusp BIC.
		(b) Difference $|a_{4,-2}|-|a_{4,4}|$ along the family
		parameterized by $\eta L=h-h_*$ with $r=r(\eta)$.
		(c) Unwrapped polarization angle near the critical direction
		$\theta_c=0$. The black curve shows $\phi_4/\pi$, and the
		colored curves show $\phi/\pi$ of nearby resonant states for
		$\delta/2\pi=0.01$ and $0.02$. Since ${\bm d}_4(0)=0$,
		$\phi_4(0)$ is undefined.
		The polarization varies over an angular scale
		$|\theta-\theta_c|\sim\delta^2$ indicated by the dashed lines.
		(d) Pitchfork bifurcation of BICs along ${\bm p}(\eta)$.
		The black and orange curves represent the at-$\Gamma$ and off-$\Gamma$
		BICs, respectively.
	}
	\label{fig:c6v_transition}
\end{figure}

We illustrate this mechanism using a family of {\em B}$_2$ BICs with
$\mu=4$ in the same $C_{6v}$ PhC slab. 
A $\mu=4$ BIC has codimension one, 
requiring only one independent condition to suppress the lower-order radiation. 
We therefore take
${\bm p}=(h,r)$, and the resulting curve of BICs is
shown in Fig.~\ref{fig:c6v_transition}(a). Along this family, we identify
a codimension-two transition point satisfying
$|a_{4,-2}|=|a_{4,4}|$, located at
${\bm p}_*=(0.954,0.357)L$ with $\omega_*L/2\pi=0.713$.
We parameterize the BIC family by 
$\eta L=h-h_*$ and $r=r(\eta)$. Figure~\ref{fig:c6v_transition}(a) and (b)
show that the transition occurs precisely at
$\eta=0$, separating the $q=-2$ and $q=4$
regions.
In the vicinity of this BIC curve, the surrounding parameter region is
populated by ordinary $\mu=2$ BICs with $q=-2$, whereas the $q=4$ BICs
form a line that terminates at the transition point, which we refer to
as the cusp BIC. 
At this point, further calculations give
$a_{4,-2}=-a_{4,4}$, so the zero set of ${\bm d}_\mu(\theta)$ is
${\cal Z}_4=\{\theta_c=j\pi/3:j=-3,\ldots,2\}$.
For $\theta_c\in{\cal Z}_4$,
${\bm d}_5(\theta_c)=0$, while
${\bm d}_6(\theta_c)\ne0$ and ${\bm b}\perp{\bm c}$.
Consequently, along these directions,
$Q\sim1/\delta^{12}$ instead of $Q\sim1/\delta^8$, and the sharp
polarization variation near $\theta_c=0$ is shown in
Fig.~\ref{fig:c6v_transition}(c).

\begin{figure}[htbp]
	\centering
	\includegraphics[scale=0.33]
	{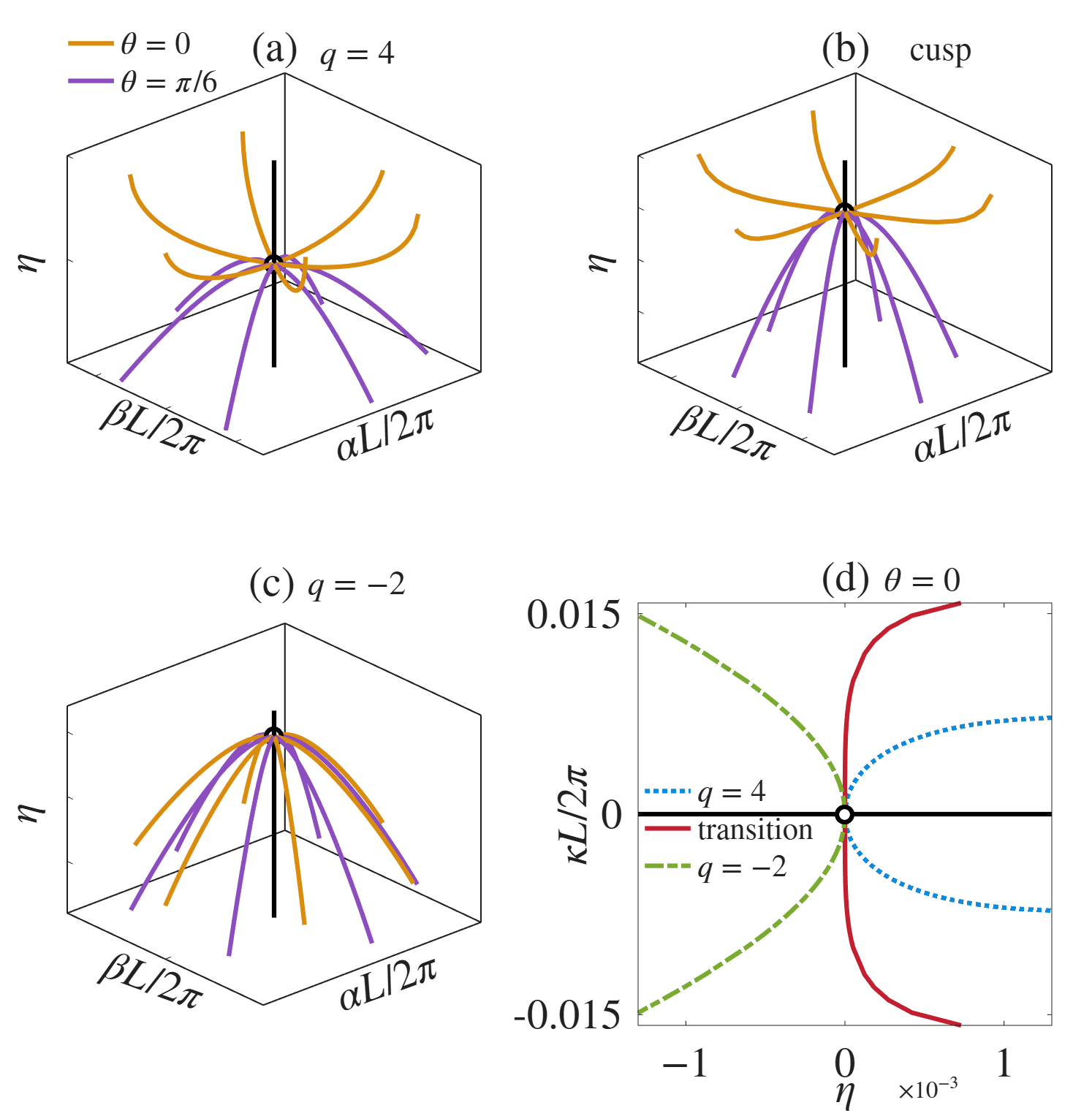}
	\caption{
		Bifurcation of off-$\Gamma$ BICs for fixed-$r$ structural
		paths across the topological transition.
		(a)--(c) BIC trajectories in $({\bm\kappa},\eta)$ space.
		The blue branches form the $\theta=0$ sextet, while the red branches
		form the $\theta=\pi/6$ sextet.
		The black lines represent the at-$\Gamma$ BIC branches.
		For $q=4$, the bifurcating sextet switches from one angular family to
		the other as $\eta$ changes sign, whereas for $q=-2$ the two sextets
		bifurcate on the same side of $\eta=0$.
		At the cusp BIC, the $\theta=0$ sextet becomes critical.
		(d) Radial positions of the bifurcating BICs along
		$\theta=0$ as functions of $\eta$.
		The $q=-2$ and $q=4$ cases follow
		$\kappa\sim|\eta|^{1/2}$, whereas the cusp BIC follows
		$\kappa\sim|\eta|^{1/4}$.
		Negative values of $\kappa$ are used to
		represent the opposite direction $\theta=-\pi$.
	}
	\label{fig:bifurcation}
\end{figure}

The zeros in ${\cal Z}_4$ govern the bifurcation of BICs along the family
${\bm p}(\eta)$.
Six symmetry-related BIC branches emerge with nonzero wavevectors
${\bm \kappa}_{\rm BIC}=\kappa{\bm \theta}_c$,
where $\theta_c\in{\cal Z}_4$. 
These branches are collectively referred
to as the $\theta=0$ sextet.
Here, mirror symmetry gives ${\bm a}\parallel{\bm c}$ in
Eq.~\eqref{eq:fixed_mu_local}, and their relative sign determines the
side of $\eta=0$ on which the bifurcation occurs, while
$\kappa\sim|\eta|^{1/2}$.
Together with the persistent at-$\Gamma$ BIC branch, the $\theta=0$
sextet forms the pitchfork bifurcation shown in
Fig.~\ref{fig:c6v_transition}(d).

We next consider the second type, in which the leading radiation order
increases from $\ell$ to $\mu=\ell+p_1$ at $\eta=0$. For a generic
structural path,
${\bm d}_\ell(\theta;\eta)\sim\eta{\bm a}_\ell(\theta)$, giving
\begin{equation}
	{\bm d}({\bm\kappa};\eta)
	\sim
	\delta^\ell
	\left[
	\eta{\bm a}_\ell(\theta)
	+
	\delta^{p_1}{\bm d}_\mu(\theta;0)
	\right].
	\label{eq:changing_mu_local}
\end{equation}
Nearby off-$\Gamma$ BICs bifurcating from the critical point also satisfy the
leading-order vanishing condition.
If ${\bm d}_\mu$ further vanishes along a critical direction $\theta_c$,
the first nonzero higher-order radiation term must also be retained.
This can give rise to a higher-order bifurcation along the critical direction.

We illustrate this case using three structural paths with fixed $r$ and
varying $h$, ${\bm p}(\eta)=(h_*+\eta L,r_*)$, through representative
$\mu=4$ BICs: a $q=-2$ BIC at
${\bm p}_*=(0.940,0.350)L$, the cusp BIC, and a $q=4$ BIC at
${\bm p}_*=(0.956,0.358)L$.
The bifurcation behavior
of the $q=-2$ BIC has been reported previously~\cite{Kang22LSA},
whereas, to our knowledge, the $q=4$ and cusp cases have not been
analyzed.
As shown in Fig.~\ref{fig:bifurcation}, the bifurcation pattern changes
across the topological cusp. For $q=-2$, the $\theta=0$ and
$\theta=\pi/6$ sextets bifurcate on the same side of $\eta=0$. For
$q=4$, the bifurcating sextet switches from one angular family to the
other as $\eta$ changes sign. In both cases,
$\kappa\sim|\eta|^{1/2}$. At the cusp BIC, the
$\theta=\pi/6$ sextet retains this scaling. For the $\theta=0$
sextet, ${\bm d}_4$ and ${\bm d}_5$ vanish along the critical
directions, while ${\bm d}_6$ remains nonzero. The bifurcation is
therefore fourth order, giving $\kappa\sim|\eta|^{1/4}$. The cusp BIC
thus marks both a change in the bifurcation pattern and the emergence
of a higher-order bifurcation.

The two examples also demonstrate the path dependence of the local
bifurcation at the same cusp BIC. Along the $\mu=4$ family, only the
$\theta=0$ sextet bifurcates. Along a structural path with fixed $r$,
the $\theta=\pi/6$ sextet also appears. The two sextets exhibit
distinct local bifurcation scalings. This path dependence reflects the
local geometry of the BIC set in ${\cal P}\times{\rm BZ}$.

In summary, we have established a leading-order angular description of
BIC topology. By isolating the leading radiation ${\bm d}_\mu$, the
topological problem is reduced to a finite set of angular harmonics.
This differs fundamentally from symmetry analyses of nearby resonances,
which generally contain infinitely many symmetry-allowed harmonics,
including higher-order contributions that do not determine the BIC
topology in the asymptotic limit~\cite{Zhen14PRL,jones2026symmetry}.
Since ${\bm d}_\mu$ can be obtained directly from the BIC, this
formulation also provides a practical route for identifying high-charge
BICs and desired polarization structures.

Our analysis further shows that zeros of the leading radiation mark
topological transitions and the bifurcation of BICs. Isolated angular
zeros give rise to sharp polarization variations and distinct
bifurcation behaviors. Our theory
also clarifies the relation between the topological charge and the
asymptotic $Q$-factor scaling: the former is determined by the angular
structure of ${\bm d}_\mu$, whereas the latter is set solely by $\mu$.
The same framework can be naturally extended to
topological structures involving circular polarization. 
The topological leading angular structure therefore provides a new tool for
exploring and engineering BIC topology, with potential applications in
structured light, topological photonics, and resonant photonics.

\bibliography{main}
\end{document}